\documentclass[twocolumn]{aastex701}

\graphicspath{{./}{figures/}}
\usepackage[position = top]{subfig}
\usepackage{amsmath}

\usepackage{appendix}
\usepackage{graphicx}
\usepackage{mathtools}
\usepackage{diagbox}
\usepackage{footmisc}
\usepackage{multirow}
\usepackage{tcolorbox}
\usepackage{physics}
\usepackage{booktabs}
\usepackage{tikz}
\usepackage{bm}
\usepackage{amssymb}
\usepackage{array}
\usepackage{xcolor}
\usepackage{soul}
\usepackage{ulem}
\usepackage{enumitem}
\usepackage[dvipsnames]{xcolor}
\usepackage[scr=boondoxo,cal=zapfc,calscaled=1.10]{mathalfa}

\usepackage{xcolor}

\definecolor{darkgreen}{RGB}{0, 100, 0}

\begin{document}

\title{Can magnetic reconnection power neutrino emission from AGN coronae?}

\author[0000-0001-6155-2827]{Omar French}
\affiliation{Center for Integrated Plasma Studies, Department of Physics, 390 UCB, University of Colorado, Boulder, CO 80309-0390, USA}
\email{omar.french@colorado.edu}

\author[0000-0001-9039-9032]{Gregory R. Werner}
\affiliation{Center for Integrated Plasma Studies, Department of Physics, 390 UCB, University of Colorado, Boulder, CO 80309-0390, USA}
\email{greg.werner@colorado.edu}

\author[0000-0003-0936-8488]{Mitchell C. Begelman}
\affiliation{JILA, University of Colorado and National Institute of Standards and Technology, 440 UCB, Boulder, CO 80309-0440, USA}
\affiliation{Department of Astrophysical and Planetary Sciences, 391 UCB, Boulder, CO 80309-0391, USA}
\email{mitch@jila.colorado.edu}

\begin{abstract}

We investigate whether reconnection of small-scale current sheets in transrelativistic supermassive black hole (SMBH) coronae can supply the nonthermal protons needed for high-energy neutrino emission, using NGC~1068 as a test case. We model the corona as a strongly turbulent, low-$\beta$, collisionless hydrogen plasma with characteristic size~$r_{\rm co}$, magnetic field strength~$B$, proton density~$n_p$, and radiation energy density~$u_{\rm rad}$. Combining the observed IceCube-band neutrino luminosity with the X-ray luminosity and Thomson optical depth reduces these coronal quantities to a one-parameter family. Across this family, the proton magnetization $\sigma_p \equiv B^2/(4\pi n_p m_p c^2)$ is transrelativistic with $\sigma_p \sim 0.3$. In this regime, we show that repeated encounters with intermittent reconnecting current sheets can energize suprathermal protons up to tens of PeV before photomeson cooling limits further acceleration. These injected particles may then be further processed by stochastic interactions with the turbulent cascade. Motivated by PIC simulations of strong turbulence at comparable magnetization, we adopt a nonthermal proton spectrum with an independently specified index and find that the predicted TeV spectral shape is broadly consistent with NGC~1068 without fitting the proton spectral slope.

\end{abstract}


\section{Introduction} \label{sec:intro}

The IceCube Neutrino Observatory has detected a growing number of high-energy neutrino excesses associated with supermassive black holes (SMBHs) in active galactic nuclei (AGN), including transient associations with blazar flares \citep{IceCube2018} and a persistent excess from the Seyfert galaxy NGC~1068 \citep{IceCube2020}. The benchmark individual source remains the~$4.2$-standard-deviation excess of~$\sim 79$ neutrino events with energies~$\sim 1.5$--$15$~TeV observed from the direction of the nearby Type~II Seyfert galaxy NGC~1068 over ten years of IceCube operation \citep{IceCube2022}. Subsequent IceCube searches find that NGC~1068 remains the most significant northern-sky neutrino source among preselected candidates and report stacked evidence from X-ray-bright Seyfert populations in both hemispheres \citep{IceCube2026XrayAGN,IceCube2026SouthernSeyferts}. The absence of accompanying TeV $\gamma$-rays suggests that the neutrinos originate from a compact, $\gamma$-ray-opaque region such as the SMBH corona \citep{Murase2020b,Inoue2020}.

Nearly five decades ago, AGN cores were proposed as powerful neutrino sources \citep{Eichler1979}. While there exists a menagerie of hadronic reactions that can produce neutrinos in principle, the most common reactions are the proton-proton ($pp$) and proton-photon ($p\gamma$) collisions that produce charged pions, owing to the abundance of reactants and their relatively large inelastic scattering cross sections. For the resulting neutrinos to possess~$\varepsilon_\nu \sim 1.5$--$15$~TeV energies (as detected in the direction of NGC~1068), the reactant protons must possess energies of~$\gamma_p m_p c^2 \sim 30$--$300$~TeV. Therefore, the operating nonthermal particle acceleration (NTPA) mechanism(s) must be capable of producing such protons despite electromagnetic and hadronic losses.

Any hadronic process that produces neutrinos through charged pion decay simultaneously produces $\gamma$-rays through neutral pion decay ($\pi^0 \to 2\gamma$). The MAGIC telescope found no TeV $\gamma$-ray emission from NGC~1068 \citep{MAGIC2019}. This non-detection requires the neutrino source to be opaque to~$\gamma$-rays. The compact and radiation-dense SMBH corona satisfies this requirement \citep{Murase2020b,Inoue2020}.

The SMBH corona is a compact region of magnetized plasma that spans a few to tens of gravitational radii from the black hole. Hot coronal electrons up-scatter thermal UV and soft X-ray photons from the accretion disk into the characteristic hard X-ray power-law spectrum observed in Seyfert galaxies \citep{Sunyaev1980}. The corona has moderate Thomson optical depth $\tau_T \sim 1$ and is sustained as a turbulent medium by the magnetorotational instability (MRI) and dynamo action. The cold proton magnetization $\sigma_p \equiv B^2/(4\pi n_p m_p c^2)$ is expected to be at least mildly relativistic ($\sigma_p \gtrsim 0.1$) so that magnetic energy is dynamically important and efficient NTPA becomes possible.

The three primary collective plasma processes for powering astrophysical NTPA are magnetic reconnection, collisionless shocks, and turbulence. In the relativistic regime, magnetic reconnection is fast and efficient at accelerating nonthermal particles, as shown through analytical methods and fully kinetic particle-in-cell (PIC) simulations \citep{Zenitani2001,Sironi2014,Guo2014,Werner2016,French2023,French2026} \citep[see][for a review]{Guo2020}. All three mechanisms may operate in the turbulent coronae of SMBHs, and a growing body of work has modeled their neutrino output. Shock-driven coronal models have been explored by \citet{Inoue2020} and \citet{Ly2026}; stochastic acceleration by turbulence \citep{Comisso2019,Zhdankin2018,Wong2020,Demidem2020,Wong2025} has been studied by \citet{Murase2020b}, \citet{Fiorillo2024b}, and \citet{Lemoine2025}; and reconnection-powered models have been developed by \citet{Fiorillo2024}, \citet{Karavola2025a}, and \citet{Mbarek2024}. These studies have shown that coronal environments can plausibly produce neutrino fluxes consistent with IceCube observations.

However, the principal obstacle to modeling coronal neutrino emission is constraining the coronal environment itself, as it sets the mechanisms and rates of proton acceleration, cooling, and escape, which together determine the resulting neutrino spectral energy distribution. Prior studies have typically chosen the proton spectral slope to fit the observed neutrino spectrum.

To address this challenge, we adopt a simple single-zone corona described by four parameters $(B, n_p, u_{\rm rad}, r_{\rm co})$ and use the IceCube-band neutrino luminosity, X-ray luminosity, and an adopted Thomson depth as constraining inputs. These inputs reduce the coronal model to a one-parameter family for NGC~1068. For each member of this family, we compute the encounter-limited acceleration of protons by intermittent reconnecting current sheets and determine the maximum proton energy from the competition between acceleration and cooling. We then calculate the resulting neutrino SED using a fiducial proton injection index motivated by PIC simulations of transrelativistic turbulence.

The rest of this paper is organized as follows. Section~\ref{sec:conditions} presents our model and the constraining inputs for the coronal parameters. Section~\ref{sec:results} applies the model to NGC~1068 and compares the resulting SED with IceCube observations. Section~\ref{sec:conclusions} states our conclusions.

\section{Neutrinos from a reconnection-powered corona} \label{sec:conditions}

The rate of neutrino production is set by the \textit{neutrino emissivity}
\begin{equation} \label{eq:neutrino_emissivity}
    \varepsilon_\nu j_{\varepsilon_\nu} = m_p c^2 \, \tau_\nu^{-1}(\gamma_p) \, \gamma_p^2 \, f_p(\gamma_p),
\end{equation}
where~$j_{\varepsilon_\nu}$ is the rate at which kinetic energy is radiated in neutrinos per unit volume, per unit neutrino energy~$\varepsilon_\nu \simeq 0.05\,\gamma_p m_p c^2$, and~$\varepsilon_\nu j_{\varepsilon_\nu}$ is the same quantity per unit logarithmic neutrino energy. The total emissivity is~$\int j_{\varepsilon_\nu}\,d\varepsilon_\nu = \int \varepsilon_\nu j_{\varepsilon_\nu}\,d\ln\varepsilon_\nu$. Here~$f_p(\gamma_p)$ is the steady-state spectrum of accelerated protons, with its absolute normalization set by the proton energy budget below. The factor~$\gamma_p^2$ combines the proton kinetic energy~$\gamma_p m_p c^2$ with the Jacobian~$d\gamma_p/d\ln\varepsilon_\nu = \gamma_p$ that converts from~$d\gamma_p$ to the logarithmic SED form. The quantity~$\tau_\nu^{-1}$ is the rate at which proton energy is channeled into neutrinos,
\begin{equation} \label{eq:tau_nu}
    \tau_\nu^{-1}(\gamma_p) \equiv \tfrac{1}{2} \, \tau_{\rm cool,\,pp}^{-1} + \tfrac{3}{8} \, \tau_{\rm cool,\,pm}^{-1},
\end{equation}
where~$\tau_{\rm cool,\,pp}^{-1}$ and~$\tau_{\rm cool,\,pm}^{-1}$ are the cooling rates due to inelastic proton-proton collisions and photomeson reactions, respectively (cf. Appendix~\ref{ss:cooling_rate}). The coefficients~$1/2$ and~$3/8$ account for the fraction of pion energy channeled to neutrinos: roughly~$3/4$ of charged pion energy goes to neutrinos via the decay chain~$\pi^+ \to \mu^+ \nu_\mu \to e^+ \bar{\nu}_\mu \nu_e \nu_\mu$ (and charge-conjugate chain for~$\pi^-$), while~$\sim 2/3$ of pions from~$pp$ collisions are charged versus~$\sim 1/2$ from photomeson reactions, giving~$\tfrac{3}{4} \times \tfrac{2}{3} = \tfrac{1}{2}$ and~$\tfrac{3}{4} \times \tfrac{1}{2} = \tfrac{3}{8}$, respectively.

\begin{figure*}[htp!]
    \centering
    \includegraphics[width=\linewidth]{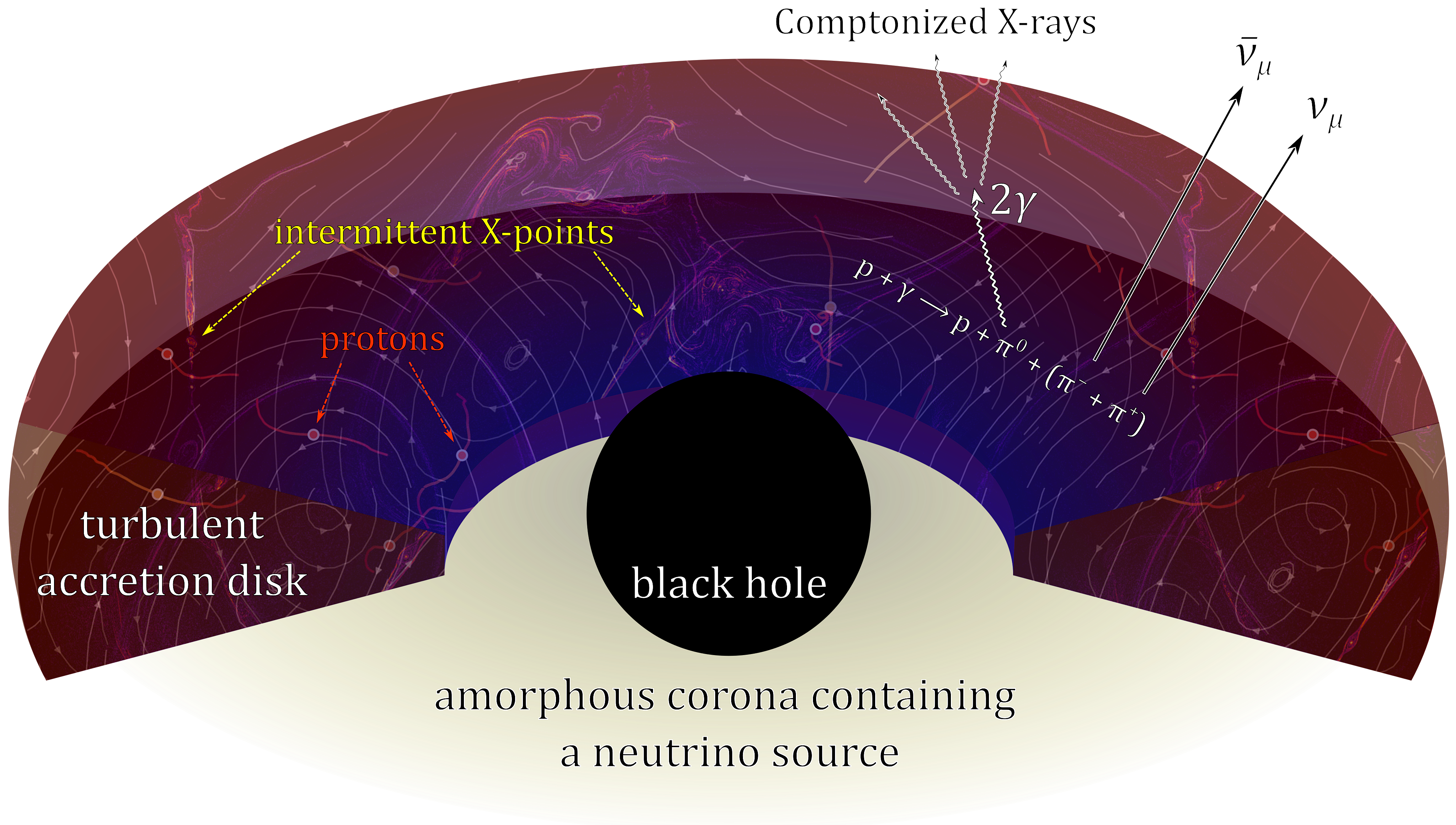}
    \caption{Cartoon of a turbulent black hole corona hosting a neutrino source. Protons are steadily energized by parallel electric fields within the reconnection diffusion regions they encounter intermittently. Inelastic collisions with ambient photons ($p\gamma$) and protons ($pp$) produce charged pions whose decays yield neutrinos, and neutral pions whose decays yield $\gamma$-rays that are absorbed by $\gamma\gamma$ pair production and reprocessed through electromagnetic cascades to lower energies within the optically thick corona.
    }
    \label{fig:intermittent_corona}
\end{figure*}

The steady-state proton spectrum~$f_p$ is determined by a transport equation that balances injection, radiative cooling, and escape,
\begin{equation} \label{eq:transport}
    \frac{\partial}{\partial \gamma_p} \left[ \dot{\gamma}_p \, f_p \right] = Q(\gamma_p) - \frac{f_p}{\tau_{\rm esc}},
\end{equation}
where~$Q(\gamma_p)$ is the proton injection spectrum and~$\tau_{\rm esc}$ is the escape timescale (both specified by the coronal model in Section~\ref{ss:model}). The rate at which~$\gamma_p$ decreases due to cooling is~$\dot{\gamma}_p \equiv d\gamma_p/dt = -\gamma_p \, \tau_{\rm cool}^{-1}$, where~$\tau_{\rm cool}^{-1} \equiv \tau_{\rm cool,\,sy}^{-1} + \tau_{\rm cool,\,ic}^{-1} + \tau_{\rm cool,\,pp}^{-1} + \tau_{\rm cool,\,pm}^{-1} + \tau_{\rm cool,\,bh}^{-1}$ is the aggregate cooling rate (Appendix~\ref{ss:cooling_rate}).\footnote{Equation~\eqref{eq:transport} treats all energy losses as continuous, even though~$pp$ and photomeson reactions are catastrophic in that a proton loses a fraction~$\kappa_{\rm pp} \sim 0.5$ or~$\kappa_{\rm pm} \sim 0.2$ of its energy per collision (Appendix~\ref{ss:cooling_rate}). This continuous-loss approximation preserves the power-law spectral index and mainly affects the normalization of~$f_p$.} When cooling dominates over escape ($\tau_{\rm cool} \ll \tau_{\rm esc}$), the steady-state solution for a power-law injection~$Q \sim \gamma_p^{-s}$ is
\begin{equation} \label{eq:steepening}
    f_p(\gamma_p) = \tau_{\rm cool}(\gamma_p) \, \gamma_p^{-1} \int_{\gamma_p}^{\gamma_{p,\rm max}} Q(\gamma_p')\,d\gamma_p'.
\end{equation}
When cooling is dominated by synchrotron or inverse-Compton in the Thomson regime, the spectrum is steepened by one power, and when dominated by photomeson reactions or Bethe-Heitler pair production in a photon spectrum~$dn_\gamma/d\varepsilon_\gamma \sim \varepsilon_\gamma^{-\alpha}$, the spectrum is steepened by~$\alpha - 1$. In the opposing limit where escape dominates, $f_p \approx \tau_{\rm esc} \, Q$.

The cooling rates are determined by the local magnetic field strength~$B$, proton density~$n_p$, and photon spectrum~$dn_\gamma/d\varepsilon_\gamma$ (Appendix~\ref{ss:cooling_rate}). The proton injection spectrum~$Q(\gamma_p)$ and escape rate~$\tau_{\rm esc}^{-1}$, by contrast, depend on the larger-scale dynamics of the source and so require modeling.

\subsection{Model of a reconnection-powered corona} \label{ss:model}

We consider a corona of size~$r_{\rm co}$ composed of a low~$\beta$, collisionless, hydrogen (proton-electron) plasma, illustrated in Figure~\ref{fig:intermittent_corona}. The plasma supports strong Alfv\'enic turbulence with dynamic alignment ($\delta B \sim B$ at the driving scale~$\ell$) wherein the cascade of perpendicular magnetic fluctuations obeys~$\delta B(k_\perp) \sim k_\perp^{-1/4}$ \citep{Boldyrev2006,Mallet2017}. Following the tearing-mediated cascade picture, we take the magnetohydrodynamic (MHD) cascade to transition near the proton skin depth~$d_p \equiv c/\omega_{\rm pp}$, where tearing modes can grow faster than eddy turnovers \citep{Loureiro2017,LoureiroBoldyrev2017} (where~$\omega_{\rm pp} \equiv \sqrt{4\pi n_p e^2/m_p}$ is the proton plasma frequency).
This produces small-scale reconnecting current sheets in which protons can be energized.

We now estimate the rate of proton acceleration within reconnecting current sheets. In what follows, we will consider protons of gyroradius~$\rho_p(\gamma_p) \equiv \gamma_p \, m_p c^2/(eB)$ within the range~$L_{\rm min} \lesssim \rho_p(\gamma_p) \ll \ell$, where we take~$L_{\rm min}\simeq 10d_p$ as the length of the shortest current sheet.

Due to the turbulent cascade, the magnetic field component which reconnects at a sheet of length~$L$ has a strength~$\delta B(L) \sim B \, (L/\ell)^{1/4}$. Since the guide magnetic field is the large-scale field~$B$, the guide field is strong with~$b_g \simeq (\ell/L)^{1/4} \gtrsim 1$. The reconnection inflow speed of each sheet is~$\beta_{\rm rec} \, \delta v_A(L)$, where~$\beta_{\rm rec} \simeq 0.1$ is the dimensionless reconnection rate in collisionless plasmas \citep{Goodbred2022,Liu2025} and~$\delta v_A (L)$ is the Alfv\'en speed of magnetic fluctuations of scale~$L$. We assume the latter to be nonrelativistic, and thus~$\delta v_A(L) \sim v_A \, (L/\ell)^{1/4}$.

A strong guide field can suppress acceleration mechanisms which operate outside the diffusion region, such as Fermi reflections off freshly-reconnecting field lines \citep{French2023}, and can stabilize the diffusion region \citep{Dahlin2017}. We therefore model the proton energy gain as direct energization by the reconnection electric field
\begin{equation} \label{eq:E_rec}
    E_{\rm rec}(L) \simeq \delta B(L) \, \frac{\delta v_A(L)}{c} \, \beta_{\rm rec} \sim \left( \frac{\ell}{L} \right)^{-1/2} \, B \, \beta_A \, \beta_{\rm rec}, 
\end{equation}
where~$\beta_A \equiv v_A/c$ is the bulk Alfv\'en speed normalized to~$c$. This results in an acceleration rate of
\begin{equation} \label{eq:gam_acc}
    \dot\gamma_{\rm acc}(L) \simeq \beta_{\rm rec} \, \beta_A \, \omega_{\rm cp} \, \left( \frac{L}{\ell} \right)^{1/2},
\end{equation}
where~$\omega_{\rm cp} \equiv eB/(m_pc)$ is the proton gyrofrequency. However, for a proton of Lorentz factor~$\gamma_p$, only sheets of length exceeding its gyroradius (i.e.,~$L\gtrsim \rho_p(\gamma_p) \equiv \gamma_p \, c/\omega_{\rm cp}$) can contribute to its acceleration. The acceleration rate of Eq.~\eqref{eq:gam_acc} must be modified to take this into account.

Geometrically, let us consider the sheets as thin folded ribbons of thickness~$\sim 2 d_p$ across the reconnection magnetic field, length~$L$ along the reconnecting field, and depth~$w \sim L \, b_g \simeq L^{3/4} \, \ell^{1/4}$ along the guide-field direction \citep{Bateman1978,Serrano2025,Davis2026}. As a result, each current sheet of length~$L$ possesses a cross-sectional area
\begin{equation} \label{eq:a_cs}
    A_{\rm cs}(L) \simeq L^{7/4} \, \ell^{1/4}.
\end{equation}
Let us further assume that current sheets are distributed according to a power-law distribution, i.e.
\begin{equation}
    \frac{dN_{\rm cs}}{dL} \sim L^{-x},
\end{equation}
as has been suggested by resistive MHD \citep{Zhdankin2016} and particle-in-cell \citep{Serrano2025,Davis2026} simulations.

We adopt~$x=7/2$ as a fiducial value for two reasons. First, with the above geometric scalings, the power dissipated by one sheet scales as
\begin{equation}
    P_{\rm cs}(L) \sim \frac{\delta B(L)^2}{8\pi} \, \delta v_A(L) \, \beta_{\rm rec} \, A_{\rm cs}(L) \sim L^{5/2}.
\end{equation}
Therefore the dissipated power per logarithmic interval in sheet length scales as
\begin{equation}
    \frac{dP}{d\ln L} \sim L \, P_{\rm cs}(L) \, \frac{dN_{\rm cs}}{dL} \sim L^{7/2-x}.
\end{equation}
Thus~$x=7/2$ corresponds to comparable dissipated power in each logarithmic interval of~$L$. Second, this value is close to the current-sheet length distributions of~$dN_{\rm cs}/dL\sim L^{-3.3}$ found in strong guide-field, incompressible RMHD intermittency simulations \citep{Zhdankin2016}. We therefore use~$x=7/2$ as an analytically convenient fiducial exponent, while interpreting it as representative of the steep intermittent-sheet regime.

\begin{figure*}[htp!]
    \centering
    \includegraphics[width=\linewidth]{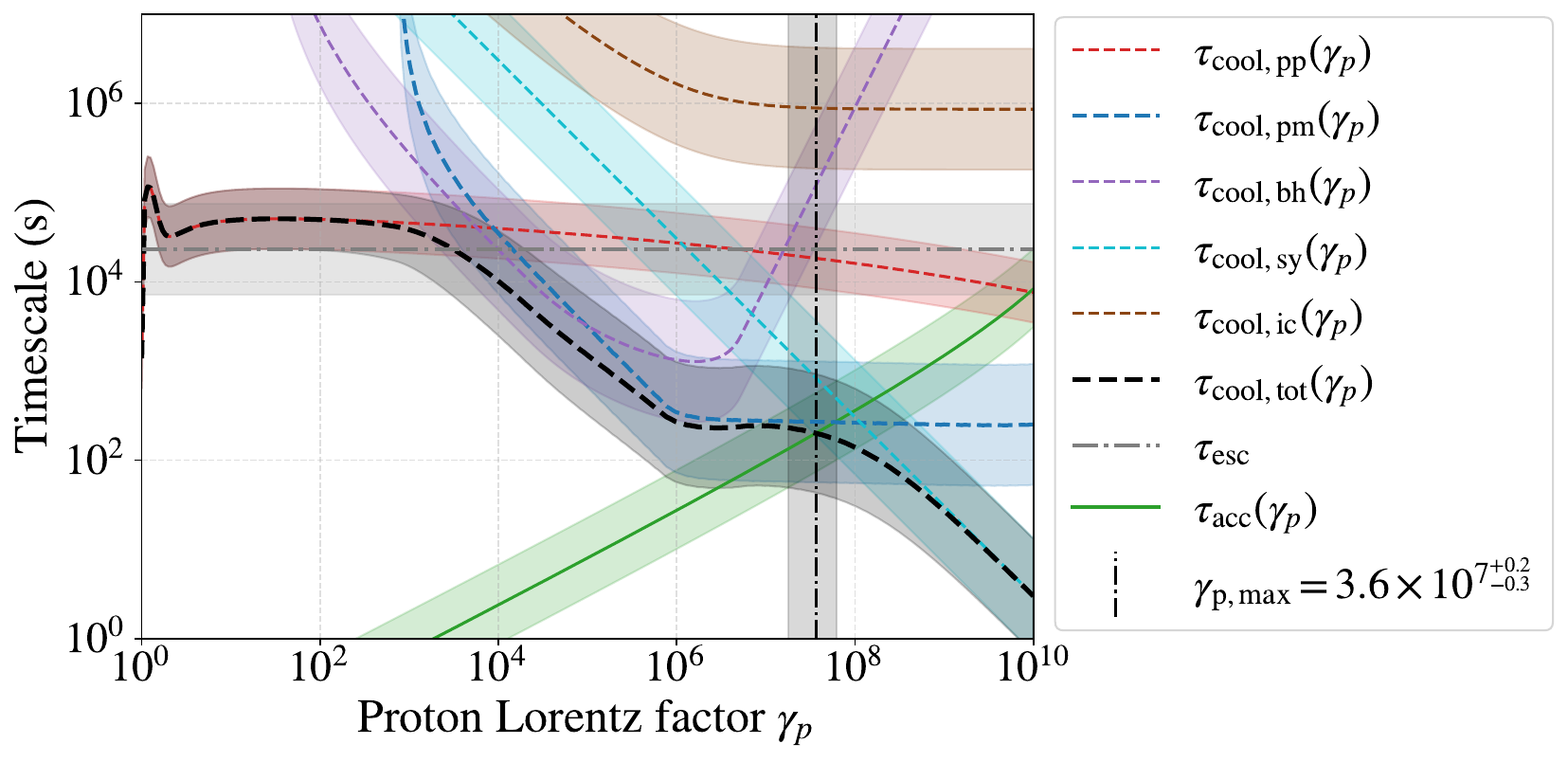}
    \caption{Phase diagram of energy-dependent acceleration and cooling timescales for NGC~1068-like SMBH corona parameters and uncertainties (cf.\ Table~\ref{tab:NGC1068_corona}). The acceleration curve~$\tau_{\rm acc}(\gamma_p)$ uses the encounter-limited current-sheet acceleration time of Eq.~\eqref{eq:tau_acc_enc}, which averages over the accessible sheet hierarchy for~$dN_{\rm cs}/dL\sim L^{-7/2}$ and includes the mean free path between accessible sheet encounters. The resulting cutoff occurs at~$\gamma_{p,\rm max}m_pc^2\sim {\rm tens}$ of PeV. The escape timescale~$\tau_{\rm esc}$ (gray dash-dot) is comparable to cooling only near the low-energy edge and exceeds the aggregate cooling timescale through most of the IceCube-producing range.}
    \label{fig:regime_phase_diagram}
\end{figure*}

Conditioned on encountering a sheet large enough to confine a proton of Lorentz factor~$\gamma_p$, the probability that the sheet has length~$L$ is therefore
\begin{equation}
    \mathcal{P}_{\rm enc}(L|\gamma_p) \, dL = \frac{L^{7/4-x} \, dL}{\int_{\rho_p(\gamma_p)}^\ell L^{7/4-x} \, dL}.
\end{equation}
The modified acceleration rate is thus
\begin{equation}
    \langle \dot\gamma_{\rm acc} \rangle_{\rm enc} = \int_{\rho_p(\gamma_p)}^\ell \dot{\gamma}_{\rm acc}(L) \, \mathcal{P}_{\rm enc}(L|\gamma_p) \,dL.
\end{equation}
For~$x=7/2$, this results in an acceleration timescale of
\begin{equation} \label{eq:tau_acc_sheet}
    \tau_{\rm acc}^{\rm sheet}(\gamma_p) \simeq \frac{1}{3} \, \beta_{\rm rec}^{-1} \, \beta_A^{-1} \, \omega_{\rm cp}^{-1} \left(\frac{\ell}{c/\omega_{\rm cp}}\right)^{1/2} \gamma_p^{1/2}.
\end{equation}
However, Eq.~\eqref{eq:tau_acc_sheet} is the acceleration time conditioned on the proton being inside an accessible current sheet. Along an actual trajectory, the acceleration rate is reduced by the duty cycle of such encounters, i.e.,
\begin{equation} \label{eq:tau_acc_enc1}
    \tau_{\rm acc}(\gamma_p) = \tau_{\rm acc}^{\rm sheet}(\gamma_p) \left[1 + \frac{\lambda_{\rm mfp}}{\left<w(L)\right>_{\rm enc}}\right],
\end{equation}
where~$\lambda_{\rm mfp}(\gamma_p)$ is the mean-free path between successive accessible sheet encounters and~$\left<w(L)\right>_{\rm enc}$ is the mean residence length in an accessible sheet. These are given by
\begin{equation} \label{eq:mfp1}
    \lambda_{\rm mfp}(\gamma_p) \simeq \frac{4d_p}{\xi} \left( \frac{\rho_p(\gamma_p)}{L_{\rm min}} \right)^{3/4},
\end{equation}
where~$\xi$ is the total filling fraction of current sheets in the corona.
For the cascade-balanced~$x=7/2$ hierarchy, the filling fraction entering Eq.~\eqref{eq:mfp1} is fixed by requiring the sheet population to dissipate the outer-scale cascade power. This gives
\begin{equation}
    \xi \simeq
    \frac{4}{3\beta_{\rm rec}}
    \frac{d_p}{L_{\rm min}}
    \frac{(L_{\rm min}/\ell)^{1/4}-L_{\rm min}/\ell}
    {\ln(\ell/L_{\rm min})}.
\end{equation}
For NGC~1068 this yields~$\xi \sim 10^{-4}$ for~$L_{\rm min}=10d_p$.
The mean residence length in an accessible sheet is
\begin{equation} \label{eq:wbar1}
    \left<w(L)\right>_{\rm enc} \simeq \frac{3}{4} \ell^{1/4} \rho_p(\gamma_p)^{3/4}
    \ln \left( \frac{\ell}{\rho_p(\gamma_p)} \right),
\end{equation}
where~$w(L) = L^{3/4} \ell^{1/4}$.

Inserting Eqs.~\eqref{eq:mfp1} and~\eqref{eq:wbar1} into Eq.~\eqref{eq:tau_acc_enc1} gives
\begin{equation} \label{eq:tau_acc_enc}
\begin{split}
    \tau_{\rm acc}(\gamma_p) &\simeq \frac{1}{3} \beta_{\rm rec}^{-1} \beta_A^{-1} \omega_{\rm cp}^{-1} \left( \frac{\ell}{c/\omega_{\rm cp}} \right)^{1/2} \gamma_p^{1/2} \\
    &\quad \times \left[ 1+ \frac{16d_p}{3\xi L_{\rm min}^{3/4}\ell^{1/4} \ln(\ell/\rho_p)} \right].
\end{split}
\end{equation}

Protons may also escape the acceleration region before losing their energy to pion production. We consider advective infall into the black hole as the primary escape mechanism, whose associated escape timescale is
\begin{equation} \label{eq:tau_esc}
    \tau_{\rm esc} = \frac{r_{\rm co}}{v_{\rm adv}} \simeq r_{\rm co} \, \frac{\sqrt{r_{\rm co}/r_g}}{\alpha_D \, c},
\end{equation}
where~$v_{\rm adv}$ is the radial inflow velocity. For thick or magnetically-elevated disks ($H/r \sim 1$) \citep{Narayan1995}, $v_{\rm adv} \simeq \alpha_D \, v_K$, where~$\alpha_D \sim 0.1$ is the viscosity parameter and~$v_K = c/\sqrt{\hat{r}}$ is the Keplerian velocity at~$\hat{r} \equiv r_{\rm co}/r_g$. Thinner disks give~$v_{\rm adv} \simeq \alpha_D (H/r)^2 v_K$ \citep{Shakura1973}, so Eq.~\eqref{eq:tau_esc} is a minimum estimate of~$\tau_{\rm esc}$.

Figure~\ref{fig:regime_phase_diagram} shows the resulting cooling, acceleration, and escape timescales for NGC~1068-like coronal parameters (Table~\ref{tab:NGC1068_corona} of Section~\ref{sec:results}). Escape is sub-dominant to cooling for~$\gamma_p > 10^4$, which sets a cooling break past which the steady-state proton spectrum is steepened according to Eq.~\eqref{eq:steepening}. The acceleration curve uses Eq.~\eqref{eq:tau_acc_enc}, with~$L_{\rm min}=10d_p$ and~$\xi$ fixed by requiring the current-sheet hierarchy to dissipate the cascade power. The proton cutoff is set by photomeson cooling at~$\gamma_{p,\rm max}m_pc^2\sim {\rm tens}$ of PeV across the viable NGC~1068 coronal parameter family.

The reconnection calculation above determines the acceleration rate and maximum attainable proton energy for a specified coronal environment. In this picture, direct energization by intermittent current sheets injects suprathermal protons into the broader turbulent accelerator, while subsequent stochastic interactions can shape the extended high-energy power law \citep{Lemoine2020,Lemoine2025}. We now introduce an energy-budget diagnostic that connects the cascade power supply to the neutrino luminosity and thereby constrains the coronal environment.

\subsection{Neutrino luminosity} \label{ss:neutrino-luminosity}

In steady state, energy balance requires the total reconnection dissipation rate to equal the cascade supply at the outer scale~$\ell$. The cascade supply is the magnetic energy~$(B^2/8\pi)\,r_{\rm co}^3$ contained in the corona times the cascade turnover rate~$v_A/\ell$. The neutrino luminosity in the corona is thus
\begin{equation} \label{eq:L_nu}
    L_\nu \sim P \, \eta_p \, \langle \eta_\nu \rangle \sim \frac{B^2}{8\pi} \, v_A \, \frac{r_{\rm co}^3}{\ell} \, \eta_p \, \langle \eta_\nu \rangle,
\end{equation}
where $\eta_p \simeq 0.5$ is the fraction of magnetic energy dissipated into protons \citep{Melzani2014,Rowan2017,Zhdankin2019} and~$\langle \eta_\nu \rangle$ is the fraction of proton energy channeled into neutrinos averaged over the proton spectrum. The latter is
\begin{equation} \label{eq:eta_nu}
    \langle \eta_\nu \rangle = 
    \frac{\int \tau_\nu^{-1}(\gamma_p) \, \gamma_p \, f_p(\gamma_p) \, d\gamma_p}{\int \left( \tau_{\rm cool}^{-1}(\gamma_p) + \tau_{\rm esc}^{-1} \right) \, \gamma_p \, f_p(\gamma_p) \, d\gamma_p}
\end{equation}
where~$\tau_\nu^{-1}$ is the neutrino channeling rate (Eq.~\eqref{eq:tau_nu}) and~$\tau_{\rm cool}^{-1} + \tau_{\rm esc}^{-1}$ is the total rate at which proton kinetic energy is removed from the corona.

Measurements of neutrino luminosity may be used with Eq.~\eqref{eq:L_nu} to constrain the power~$P = P(B, n_p, r_{\rm co})$, which in turn constrains the coronal parameters. The proton spectral index is specified by the kinetic closure described in Section~\ref{sec:results}, while the high-energy cutoff follows from the acceleration-cooling balance.

\section{Observational implications} \label{sec:results}

The benchmark NGC~1068 signal is the~$4.2$-standard-deviation excess of~$\sim 79$ events observed by IceCube over 10 years~\citep{IceCube2022}. A later 13.1-year northern-sky follow-up finds that NGC~1068 remains the most significant neutrino source among preselected candidates, with an unbroken power-law spectrum of index~$3.4\pm0.2$ \citep{IceCube2026XrayAGN}. We use this source to test our model.

Table~\ref{tab:NGC1068_obs} lists the quantities used as constraining inputs for the NGC~1068 corona. 
We adopt a coronal photon spectrum with $\alpha=2$ over $[100~{\rm eV},\,100~{\rm keV}]$ and obtain $L_X$ by logarithmically extrapolating the absorption-corrected intrinsic $2$--$10$~keV luminosity of \citet{Bauer2015}.
The all-flavor neutrino luminosity in the $1.5$--$15$~TeV band has been reported by \citet{IceCube2022} to be~$L_\nu \approx 3 \times 10^{42}$~erg~s$^{-1}$.
We set the minimum neutrino power in the constraint system to this observed band luminosity and set the cascade driving scale to the corona size, $\ell = r_{\rm co}$. The bolometric neutrino luminosity may be larger by roughly an order of magnitude, depending on the proton spectrum shape.
With $\eta_p\simeq0.5$ and the computed~$\langle\eta_\nu\rangle$, the observed band luminosity gives a minimum magnetic cascade power~$P=L_\nu/(\eta_p\langle\eta_\nu\rangle)\simeq(3.0$--$4.1)\times10^{43}$~erg~s$^{-1}$ across the viable family, or~$\simeq0.3$--$0.4\,L_X$.
Due to the Compton thickness of NGC 1068 \citep{Bauer2015,Marinucci2016}, the Thomson optical depth $\tau_T$ has not yet been directly measured. We therefore adopt~$\tau_T \simeq 0.3$ as a typical value \citep{Ricci2018}.
We neglect pair enrichment when applying the Thomson-depth constraint ($n_{e^-} \simeq n_p$), with the correction estimated in Appendix~\ref{app:pair_enrichment} and found to be sub-dominant to the $\lambda$-degeneracy spread. 

The fiducial coronal parameters satisfying these inputs are listed in Table~\ref{tab:NGC1068_corona}, where all four parameters are linked through the single degree of freedom~$\lambda$ from numerical solution of the constraint system (Table~\ref{tab:NGC1068_corona}). 
The four coronal parameters are obtained by solving
\begin{align}
    \tau_T &= \sigma_T n_p r_{\rm co},\\
    L_X &= 4\pi c u_{\rm rad} r_{\rm co}^2,\\
    L_\nu &= \eta_p \langle\eta_\nu\rangle
    \frac{B^2}{8\pi} v_A r_{\rm co}^2,
\end{align}
where we have set~$\ell=r_{\rm co}$.
We parametrize the remaining degree of freedom by~$\lambda\equiv B/B_0$, with~$B_0=6$~kG at the lower endpoint.
For each~$\lambda$, $B=\lambda B_0$ is fixed, $r_{\rm co}$ is solved from the neutrino-luminosity constraint, and then~$n_p$ and~$u_{\rm rad}$ follow from~$\tau_T$ and~$L_X$.
In the formal~$v_A \to c$ limit this reduces to the approximate scaling
\begin{align} \label{eq:lambda_scaling}
    B(\lambda) &= \lambda \, B_0, & n_p(\lambda) &= \lambda \, n_{p,0}, \\
    u_{\rm rad}(\lambda) &= \lambda^2 \, u_{\rm rad,0}, & r_{\rm co}(\lambda) &= \lambda^{-1} \, r_{\rm co,0}, \nonumber
\end{align}
We set the lower endpoint~$\lambda = 1$ by taking~$B_0 = 6$~kG from the external cascade-emission bound of \citet{Blanco2025}. This bound is model-dependent, but it gives a lower field scale because smaller coronal fields overproduce the sub-GeV $\gamma$-ray flux when the model is normalized to the IceCube neutrino luminosity. At this endpoint our photon field is still optically thick to TeV photons, consistent with the MAGIC non-detection \citep{MAGIC2019}. The upper bound~$\lambda \leq \lambda_{\rm max}$ comes from the requirement that the corona extend beyond the innermost stable circular orbit of a non-spinning black hole,~$r_{\rm co}(\lambda) \geq 6 \, r_g$. Together these give~$\lambda \in [1, \lambda_{\rm max}]$ with~$\lambda_{\rm max} \approx 4.3$.

\begin{table}[htp!]
    \centering
    \caption{Constraining inputs used for the NGC~1068 corona.}
    \label{tab:NGC1068_obs}
    \begin{tabular}{ll}
        \toprule
        Quantity & Value \\
        \midrule
        $dn_\gamma/d\varepsilon_\gamma \propto \varepsilon_\gamma^{-\alpha}$ & $\alpha = 2$, $\varepsilon_\gamma \in [100~{\rm eV},\, 100~{\rm keV}]$ \\
        $L_X = 4\pi c u_{\rm rad} r_{\rm co}^2$ & $\approx 9.5 \times 10^{43}$~erg~s$^{-1}$ \\
        $L_\nu \sim \eta_p\langle\eta_\nu\rangle (B^2/8\pi) \, v_A \, r_{\rm co}^2$ & $\gtrsim 3 \times 10^{42}$~erg~s$^{-1}$ \\
        $\tau_T \simeq \sigma_T n_p r_{\rm co}$ & $\approx 0.3$ \\
        \bottomrule
    \end{tabular}
\end{table}

\begin{table}[htp!]
    \centering
    \caption{Full range of possible NGC~1068 coronal parameters for the adopted constraining inputs (Table~\ref{tab:NGC1068_obs}). For numerical calculations, we adopt the central value~$\lambda = \lambda_{\rm max}^{1/2} \approx 2.1$. Shaded bands in Figures~\ref{fig:regime_phase_diagram} and~\ref{fig:icecube_comparison} span the full viable range~$\lambda \in [1,\, \lambda_{\rm max}]$.}
    \label{tab:NGC1068_corona}
    \begin{tabular}{lll}
        \toprule
        Parameter & $\lambda = 1$ & $\lambda \approx 4.3$ \\
        \midrule
        $B$ & $6$~kG & $26$~kG \\
        $n_p$ & $1.1 \times 10^{10}$~cm$^{-3}$ & $5.1 \times 10^{10}$~cm$^{-3}$ \\
        $u_{\rm rad}$ & $1.4 \times 10^{5}$~erg~cm$^{-3}$ & $3.2 \times 10^{6}$~erg~cm$^{-3}$ \\
        $r_{\rm co}$ & $29 \, r_g$ & $6 \, r_g$ \\
        \bottomrule
    \end{tabular}
\end{table}

For NGC~1068, we find that the coronal proton magnetization is~$\sigma_p \in [0.18,\,0.70]$ and thus transrelativistic.
\footnote{The upper-endpoint magnetic field~$B \approx 26$~kG is below a local MAD/ram-pressure estimate at the ISCO, $B_{\rm MAD}\sim \sqrt{2\dot{M}c}/(6r_g)\approx 70$~kG, for $M_{\rm BH}=10^7\,M_\odot$ \citep{Greenhill1996}, $\lambda_{\rm Edd}=0.5$, and radiative efficiency~$\eta=0.1$. It is also below the local magnetic-Eddington field~$B_{\rm Edd}=\sqrt{2L_{\rm Edd}/(c r_{\rm co}^2)}\approx33$~kG at $r_{\rm co}=6r_g$.}
Particle-in-cell simulations of ion-electron plasma undergoing strong turbulence have measured ion power laws with $s \gtrsim 3$ at $\sigma_i \in \{0.035,\,0.10,\,0.19\}$ \citep{Groselj2026}. Similar measurements have been found in ion-electron turbulence at $\sigma_p \approx 0.02$ ($v_A = 0.14\,c$), where a somewhat steeper $s \approx 3.5$ was found \citep{Comisso2022}, as well as in positron-electron plasma turbulence at~$v_A \approx 0.5\,c$, where~$s \approx 3$ was found \citep{Zhdankin2018}. 
We approximate the high-energy injection tail by
\begin{equation}
Q(\gamma_p)=Q_0
\begin{cases}
0, & \gamma_p < \gamma_{\rm inj},\\
\gamma_p^{-s}\exp(-\gamma_p/\gamma_{p,\rm max}), & \gamma_p \ge \gamma_{\rm inj},
\end{cases}
\end{equation}
with~$s=3.5$ and~$\gamma_{\rm inj}=10^4$.
We treat~$\gamma_{\rm inj}$ as the low-energy onset of the asymptotic tail rather than a sharp injection threshold. It lies below the~$\gamma_p\sim3\times10^4$--$3\times10^5$ protons that produce IceCube-band neutrinos and near the transition above which cooling dominates escape in Figure~\ref{fig:regime_phase_diagram}.
The cutoff~$\gamma_{p,\rm max}$ is set by the acceleration-cooling balance in Figure~\ref{fig:regime_phase_diagram}, and~$Q_0$ is fixed by the proton energy budget per unit volume, $\int Q(\gamma_p)\gamma_p m_pc^2\,d\gamma_p=\eta_p P/r_{\rm co}^3$.

\begin{figure}[htp!]
    \centering
    \includegraphics[width=\linewidth]{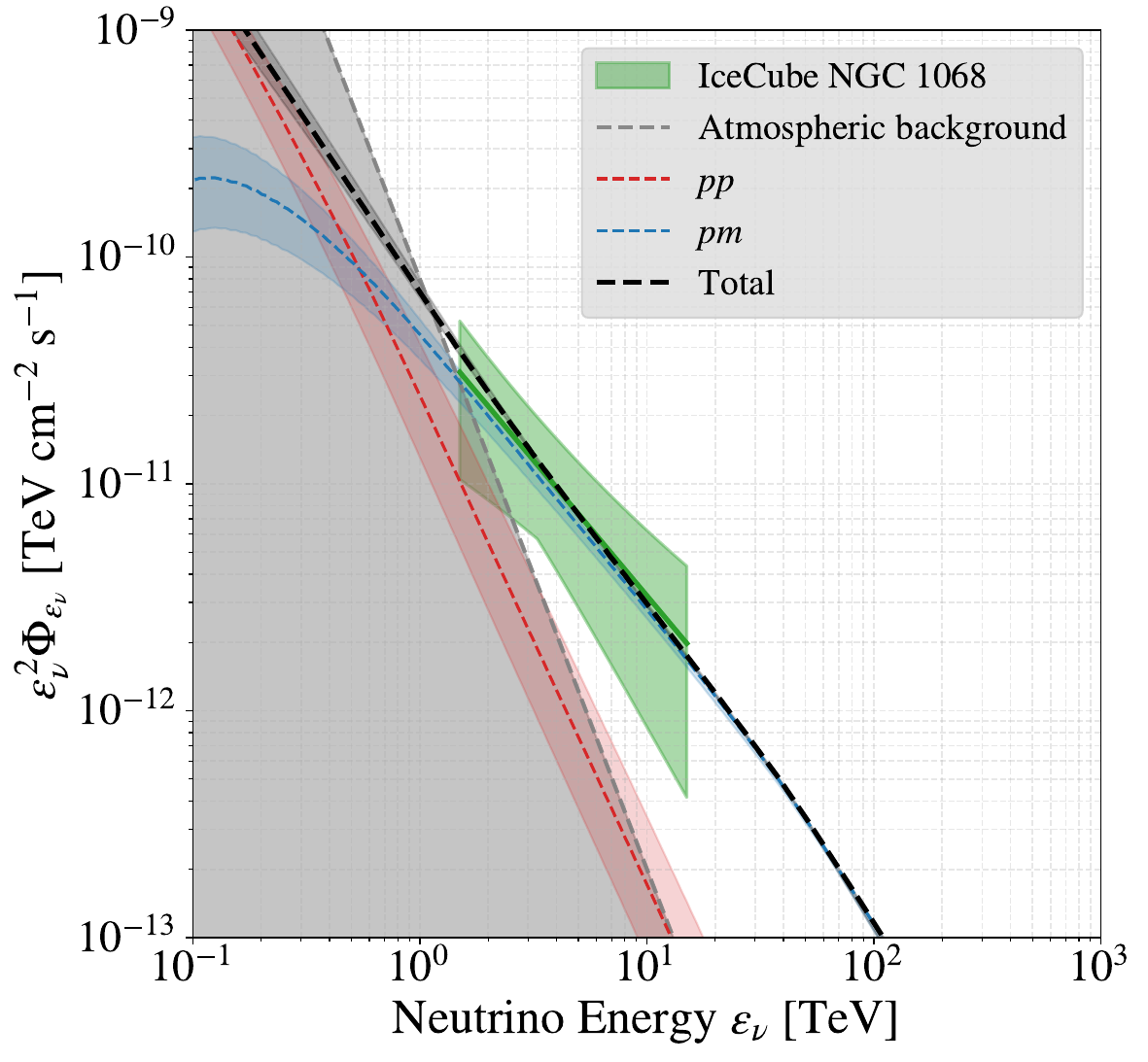}
    \caption{
    Comparison of our model neutrino SED (dashed black) with the neutrino excess detected by IceCube in the direction of NGC~1068 (green band indicates 95\% confidence). Contributions from~$pp$ (red) and~$pm$ (blue) channels are shown separately. Shaded bands show the variation across the full~$\lambda \in [1,\, \lambda_{\rm max}]$ family (Table~\ref{tab:NGC1068_corona}), with each~$\lambda$ value normalized to the IceCube flux at the band center. The gray shaded region indicates the atmospheric neutrino background, below which the source signal is indistinguishable from background.
    }
    \label{fig:icecube_comparison}
\end{figure}

The IceCube spectrum requires a steep parent proton distribution in our cooling and interaction model, and the inferred transrelativistic magnetization selects the PIC regime where such spectra are measured.\footnote{We expect the proton spectrum to harden at low~$\gamma_p \lesssim 10^4$, as also pointed out by \citet{Mbarek2024}. A pure~$\gamma_p^{-3.5}$ spectrum extending to thermal energies would place too much of the proton energy budget at low~$\gamma_p$, requiring the bolometric~$L_\nu$ to exceed the IceCube-band lower limit used in Table~\ref{tab:NGC1068_obs}.} For visualization, each model curve is normalized to the IceCube flux at the band center. The shaded bands show how the spectral shape and relative~$pp$ and~$pm$ contributions vary across the allowed~$\lambda$ family.
We compute the emissivity from Eq.~\eqref{eq:neutrino_emissivity} and, for the plotted comparison, decompose the normalized total (dashed black) into~$pp$ (red) and~$pm$ (blue) contributions.

\section{Conclusions} \label{sec:conclusions}

In this paper, we have assessed whether magnetic reconnection in a turbulent SMBH corona can power the high-energy neutrino emission observed from sources such as NGC~1068. We assume steady-state strong Alfv\'enic turbulence in a collisionless, low-$\beta_e$ corona where reconnection at kinetic-scale current sheets dissipates the cascade supply. The dissipation rate is then $P \sim (B^2/8\pi)\,v_A\,r_{\rm co}^2$ for a single-zone corona with $\ell \sim r_{\rm co}$. The observed IceCube-band luminosity supplies the minimum neutrino power in~$L_\nu = \eta_p\langle\eta_\nu\rangle P$. Together with the adopted Thomson optical depth~$\tau_T$ and X-ray luminosity~$L_X$, it reduces the four coronal parameters $(B,\,n_p,\,u_{\rm rad},\,r_{\rm co})$ to a one-parameter family~$\lambda$ (Table~\ref{tab:NGC1068_corona}). For NGC~1068, the viable range~$\lambda \in [1,\,\lambda_{\rm max} \approx 4.3]$ corresponds to $\sigma_p \in [0.18,\,0.70]$ and cooling-weighted~$\langle \eta_\nu \rangle \approx 0.15$--$0.20$.
This places NGC~1068 in the transrelativistic regime. The IceCube spectrum requires a steep parent proton distribution in our cooling and interaction model, and this magnetization selects the PIC regime where such spectra are measured \citep{Zhdankin2018,Comisso2022,Groselj2026}. Figure~\ref{fig:icecube_comparison} shows the resulting SED.

The reconnection-powered coronal models of \citet{Fiorillo2024} and \citet{Karavola2025a} consider a pair-dominated regime in which protons are sparse ($n_p \ll n_e$) and ultrarelativistically magnetized ($\sigma_p \sim 10^4$--$10^5$), producing a broken power law whose break energy~$E_{p,\rm br} \sim \sigma_p m_p c^2$ falls within or near the IceCube band and whose spectral shape depends on the assumed magnetization.
\citet{Mbarek2024} study the proton-photon-neutrino interplay directly with PIC simulations of both coronal reconnection and turbulence, though across a narrower range of proton energies than required for the IceCube band. Our model uses the same PIC-derived proton spectra and $\beta_{\rm rec} \sim 0.1$ as the prior reconnection-powered coronal works of \citet{Fiorillo2024,Karavola2025a,Mbarek2024}, and shares with the strongly turbulent coronal model of \citet{Fiorillo2024b} the assumption of transrelativistic turbulence with $\delta B \sim B$. Our cascade-supply assumption reduces the coronal parameter space to a one-parameter family with the observed IceCube-band luminosity as one of the inputs.

Our model can be refined in several ways. It adopts a single-zone corona of size~$r_{\rm co}$ with uniform~$B$, $n_p$, and~$u_{\rm rad}$. Realistic coronae possess radial gradients of these quantities that would modify the emissivity profile without necessarily altering the integrated SED. Our upper bound~$\lambda_{\rm max}$ assumes a non-spinning (Schwarzschild) black hole with innermost stable circular orbit at~$6\,r_g$. A Kerr black hole with significant prograde spin has a smaller ISCO, which could expand the viable~$\lambda$ range. Even for a Schwarzschild hole, the ISCO is not a hard boundary, as substantial dissipation can occur within the plunging region inside~$6\,r_g$, further weakening the upper bound on~$\lambda$. Our advective escape model (Eq.~\eqref{eq:tau_esc}) is energy-independent. It is comparable to cooling near $\gamma_p\sim10^4$ and subdominant at higher energies in the IceCube-producing range (Figure~\ref{fig:regime_phase_diagram}). Diffusive escape, if present, would introduce energy-dependent losses but is slower still \citep{Lemoine2025}. In the corresponding physical picture, direct reconnection electric fields inject suprathermal protons into the turbulent accelerator, and subsequent turbulent interactions can shape the high-energy spectrum \citep{Lemoine2020,Lemoine2021,Lemoine2024,Lemoine2025}. A first-principles calculation coupling reconnection injection, stochastic acceleration, turbulent escape, radiative cooling, and nonlinear damping remains open.

\begin{acknowledgments}
We thank Dmitri A. Uzdensky for helpful discussions. We acknowledge support from NASA Astrophysics Theory Program grant 80NSSC24K0941. O.F. acknowledges support from the National Science Foundation Graduate Research Fellowship under Grant No. DGE 2040434. 
\end{acknowledgments}

\appendix
\section{Timescales of proton cooling} \label{app:cooling_regimes} \label{ss:cooling_rate}

The aggregate cooling rate of protons is given by
\begin{equation} \label{eq:tau_cool_tot}
    \tau^{-1}_{\rm cool, tot}(\gamma_p) = \tau^{-1}_{\rm cool, sy} + \tau^{-1}_{\rm cool, ic} + \tau^{-1}_{\rm cool, pp} + \tau^{-1}_{\rm cool, pm} + \tau^{-1}_{\rm cool, bh},
\end{equation}
comprising synchrotron radiation, inverse-Compton (IC) scattering, proton-proton collisions, photomeson production, and the Bethe-Heitler process. We discuss each below.

The proton cooling rates from synchrotron and inverse-Compton (IC) are given by
\begin{equation} \label{eq:tau_cool_synch_IC}
    \tau_{\rm cool, sy}^{-1}(B,\gamma_p) = \frac{4}{3} \sigma_T \left(\frac{m_e}{m_p}\right)^2 c \frac{u_B}{m_p c^2} \gamma_p,
    \hspace{1cm}
    \tau_{\rm cool, ic}^{-1}(u_{\rm rad},\gamma_p) =
    \frac{4}{3}\sigma_T\left(\frac{m_e}{m_p}\right)^2 c \frac{u_{\rm rad}}{m_p c^2} \gamma_p,
\end{equation}
where~$u_B \equiv B^2/(8\pi)$ is the magnetic energy density, $B$ is the local magnetic field strength, and~$u_{\rm rad} = \int_0^\infty \varepsilon_\gamma \frac{dn_\gamma}{d\varepsilon_\gamma} \,d\varepsilon_\gamma$ is the radiation energy density. In our numerical calculations, the IC cooling rate includes an approximate Klein-Nishina suppression, which reduces the rate above~$\gamma_p \gtrsim m_p c^2 / (4\,\varepsilon_{\gamma,\rm min})$ (see Figure~\ref{fig:regime_phase_diagram}).

For the remaining cooling mechanisms, we consider a nonthermal, relativistic ($\beta_p \simeq 1$) proton immersed in a charge-neutral plasma containing various particle species. The dominant collision partners in astrophysical plasmas are thermal (low-energy) ambient protons~($i=p$) and ambient photons~($i=\gamma$). We therefore assume~$i \in \{p, \gamma\}$ and treat the number densities~$n_p, n_\gamma$ as environment-dependent inputs to our model.
In general, these number densities will depend on the energy of these species, i.e., $n_i = n_i(\varepsilon_i)$, where $\varepsilon_i$ is the kinetic energy of species~$i$ in the lab frame (e.g., the SMBH rest frame).

We denote~$\sigma_{\rm pi} = \sum_j \sigma^{(j)}_{\rm pi}$ as the total inelastic scattering cross section for a proton colliding with species~$i$. Here, $\sigma^{(j)}_{\rm pi} = \sigma^{(j)}_{\rm pi}(\gamma_p)$ is the cross section for a particular inelastic reaction~$j$ and~$\gamma_p$ is the incident proton Lorentz factor. We also denote the fraction of energy the proton loses per reaction with species~$i$ as the inelasticity~$\kappa^{(j)}_{\rm pi}(\gamma_p)$. We now treat proton-proton reactions and proton-photon reactions separately.

In proton-proton reactions, the number density is often assumed to be independent of energy \citep{Begelman1990,Murase2023}, yielding the cooling rate of the energetic proton (where~$i=p$ is retained to avoid confusing ambient protons with high-energy protons):
\begin{equation} \label{eq:tau_cool_ps}
    \tau_{\rm cool, pi}^{-1}(n_i, \gamma_p) = c \, n_i \, \sum_j \sigma^{(j)}_{\rm pi}(\gamma_p) \, \kappa^{(j)}_{\rm pi}(\gamma_p).
\end{equation}
At energies far above the pion production threshold, the proton inelasticity per~$pp$ collision is approximately~$\kappa_{\rm pp} \sim 0.5$ \citep{Sikora1987,Begelman1990,Gaisser2016}, which we adopt as a constant. Setting~$\kappa_{\rm pi}^{(j)}(\gamma_p) = \kappa_{\rm pi} = \rm const$ simplifies the~$pp$ cooling rate expression, as it also allows the total inelastic scattering cross section~$\sigma_{\rm pp}(\gamma_p)$ to be used, which is measured experimentally \citep{ParticleDataGroup2020} and can be parametrized directly \citep{Zou2024}. Hence (where again~$i=p$),
\begin{equation} \label{eq:tau_cool_ps_simplified}
    \tau_{\rm cool, pi}^{-1}(n_i, \gamma_p) = c \, n_i \, \kappa_{\rm pi} \sum_j \sigma^{(j)}_{\rm pi}(\gamma_p) = c \, n_i \, \kappa_{\rm pi} \, \sigma_{\rm pi}(\gamma_p).
\end{equation}

In proton-photon reactions, the rate of proton energy loss (assuming an isotropic photon field, ultrarelativistic protons ($\beta_p \to 1$), and the head-on approximation $1 - \beta_p \cos\theta \approx 2$) is given by \citet{Stecker1968}:
\begin{equation} \label{eq:tau_cool_pg}
    \tau_{\rm cool, p\gamma}^{-1} \left(\frac{dn_\gamma}{d\varepsilon_\gamma}, \gamma_p \right) = \frac{c}{2\gamma_p^2} \int_{\rm \bar{\varepsilon}_{\gamma, \rm th}}^\infty \left[ \bar{\varepsilon}_\gamma \, \sigma_{\rm p\gamma}(\bar{\varepsilon}_\gamma) \, \kappa_{\rm p\gamma}(\bar{\varepsilon}_\gamma) \right] d\bar{\varepsilon}_\gamma \int_{\bar{\varepsilon}_\gamma/2\gamma_p}^\infty \left[ \varepsilon_\gamma^{-2} \, \frac{d n_\gamma}{d{\varepsilon_\gamma}} \right] d\varepsilon_\gamma,
\end{equation}
where~$\varepsilon_\gamma$ is the photon energy in the lab frame, $\bar{\varepsilon}_\gamma \simeq 2\varepsilon_\gamma \gamma_p$ is the photon energy in the proton rest frame, $\bar{\varepsilon}_{\gamma, \rm th}$ is a threshold photon energy in the proton rest frame, $\sigma_{\rm p\gamma}$ is the $p\gamma$ cross section, $\kappa_{\rm p\gamma}$ is the proton inelasticity, and~$dn_\gamma/d\varepsilon_\gamma$ is the photon spectrum in the lab frame.
We treat photomeson production and the Bethe-Heitler process separately below.

For photomeson production, $\sigma_{\rm p\gamma}$ can be obtained via parameterization of experimental data from the Particle Data Group \citep{ParticleDataGroup2020,Zou2024} and the inelasticity is often assumed to be $\kappa_{\rm pm} \sim 0.2$ \citep{Kelner2008}. The threshold condition for producing pions via~$p\gamma$ interactions is given by the (Lorentz-invariant) square of the center-of-momentum energy of the interaction (i.e., the \citet{Mandelstam1958} variable):
\begin{equation} \label{eq:mandelstam}
    s_M = m_p^2 c^4 + 2 m_p c^2 \bar{\varepsilon}_{\gamma, \rm th} = \left( m_p + m_\pi \right)^2 c^4,
\end{equation}
yielding~$\bar{\varepsilon}_{\gamma, \rm th} \simeq 0.145 \, \text{GeV}$. Substituting these values into Eq.~\eqref{eq:tau_cool_pg} yields~$\tau_{\rm cool, pm}^{-1}(\gamma_p)$.

For the Bethe-Heitler process ($p\gamma \to p e^+ e^-$ \citep{Bethe1934}), if one assumes that the photon field is isotropic and that the incident proton is ultrarelativistic, then the cross section is approximately \citep{Bethe1954,Heitler1954}:
\begin{equation} \label{eq:sigma_bh}
    \sigma_{\rm bh}(\bar{\varepsilon}_\gamma) \approx \frac{7}{6}\frac{\alpha}{\pi}\,\sigma_T \left[\ln\left(\frac{2\bar{\varepsilon}_\gamma}{m_e c^2}\right) - \frac{109}{42} \right],
\end{equation}
where~$\alpha$ is the fine-structure constant. The inelasticity is approximately
\begin{equation} \label{eq:kappa_bh}
    \kappa_{\rm bh}(\bar{\varepsilon}_\gamma) \approx \frac{2 m_e}{m_p} \left[ \ln \!\left( \frac{2 \bar{\varepsilon}_\gamma}{m_e c^2} \right) - 1 \right].
\end{equation}
Typically, $\kappa_{\rm bh}(\bar{\varepsilon}_\gamma) \ll 1$ owing to~$m_e/m_p \ll 1$. Eqs.~\eqref{eq:sigma_bh} and~\eqref{eq:kappa_bh} become more accurate away from threshold, i.e., for~$2 \bar{\varepsilon}_\gamma \gg m_e c^2$. The Bethe-Heitler cooling rate~$\tau_{\rm cool,bh}^{-1}$ is obtained by inserting~$\sigma_{\rm bh}(\bar{\varepsilon}_\gamma)$ and $\kappa_{\rm bh}(\bar{\varepsilon}_\gamma)$ into Eq.~\eqref{eq:tau_cool_pg}.

The aggregate cooling rate requires knowledge of three environmental inputs. The magnetic field strength~$B$ controls~$\tau_{\rm cool, sy}^{-1}(B,\gamma_p)$, the proton density~$n_p$ controls~$\tau_{\rm cool, pp}^{-1}(n_p, \gamma_p)$, and the photon spectrum~$dn_\gamma/d\varepsilon_\gamma$ controls~$\tau_{\rm cool, ic}^{-1}(u_{\rm rad}, \gamma_p)$, $\tau_{\rm cool, pm}^{-1}(dn_\gamma/d\varepsilon_\gamma, \gamma_p)$, and~$\tau_{\rm cool, bh}^{-1}(dn_\gamma/d\varepsilon_\gamma, \gamma_p)$.

\section{Estimate of pair enrichment} \label{app:pair_enrichment}

The constraint system in Section~\ref{sec:results} infers the proton density from the Thomson optical depth using the negligible-pair approximation $n_{e^-} \simeq n_p$. Hadronic interactions also inject secondary $e^\pm$, so we check this approximation a posteriori. We define the pair-enrichment parameter
\begin{equation} \label{eq:chi}
    \chi \equiv \frac{n_{e^-} + n_{e^+}}{n_p} - 1
    = \frac{2}{n_p} \,
    \frac{\dot n_+}
    {(1 + \chi/2)\,n_p \, \pi r_e^2 c + \tau_{\rm esc}^{-1}},
\end{equation}
where $\dot n_+$ is the total positron injection rate per unit volume. The denominator includes non-relativistic annihilation on ambient electrons, with rate coefficient $\pi r_e^2 c$, and advective escape. The factor $(1+\chi/2)$ accounts for annihilation on both the original coronal electrons and the electrons created together with the positrons. For the transrelativistic solutions in Table~\ref{tab:NGC1068_corona}, fresh pairs cool rapidly before annihilating, so the non-relativistic annihilation rate is adequate.

We evaluate $\dot n_+$ using the cooled proton distribution from Section~\ref{sec:conditions}, including Bethe-Heitler pair production, $\gamma\gamma$ absorption of neutral-pion decay photons, and positrons from charged-pion decay. Across the viable $\lambda$ range, Eq.~\eqref{eq:chi} gives $\chi \simeq 0.36$--$0.37$, corresponding to $n_{e^+}/n_p \simeq 0.18$. Including this correction in the Thomson-depth constraint would lower the inferred proton density by only $(1+\chi)^{-1} \simeq 0.73$, leaving the conclusions unchanged.

\bibliography{main}{}
\bibliographystyle{aasjournal} 

\end{document}